\RequirePackage{fix-cm}
\documentclass[twocolumn]{svjour3}          % twocolumn
\smartqed  % flush right qed marks, e.g. at end of proof
\usepackage{amssymb}
\usepackage{graphicx}
\usepackage{siunitx}
\usepackage{hyperref}
\usepackage{booktabs, multicol, multirow}
\usepackage{pdflscape}
\usepackage{afterpage}
\usepackage{amsmath}
\usepackage{float}
\usepackage[caption = false]{subfig}
\begin{document}

\title{Spinal Metastases Segmentation in MR Imaging using Deep Convolutional Neural Networks}

\titlerunning{Spinal Metastases Segmentation}        % if too long for running head

\author{Georg Hille \and Johannes Steffen \and Max D\"unnwald \and Mathias Becker \and Sylvia Saalfeld \and Klaus T\"onnies}


\institute{G. Hille \at
              Department of Simulation and Graphics, University of Magdeburg, Universit\"atsplatz 2,	39106 Magdeburg, Germany\\
              Tel.: +49-391-67-52189\\
              \email{georg.hille@ovgu.de}             \\
}

\maketitle

\begin{abstract} 
\textbf{\\Purpose:} This study's objective was to segment spinal metastases in diagnostic MR images using a deep learning-based approach. Segmentation of such lesions can present a pivotal step towards enhanced therapy planning and validation, as well as intervention support during minimally invasive and image-guided surgeries like radiofrequency ablations. 
\textbf{\\Methods:} For this purpose, we used a U-Net like architecture trained with 40 clinical cases including both, lytic and sclerotic lesion types and various MR sequences. Our proposed method was evaluated with regards to various factors influencing the segmentation quality, e.g. the used MR sequences and the input dimension. We quantitatively assessed our experiments using Dice coefficients, sensitivity and specificity rates.
\textbf{\\Results:} Compared to expertly annotated lesion segmentations, the experiments yielded promising results with average Dice scores up to $\SI{77.6}{\%}$ and mean sensitivity rates up to $\SI{78.9}{\%}$.
\textbf{\\Conclusion:} To our best knowledge, our proposed study is one of the first to tackle this particular issue, which limits direct comparability with related works. In respect to similar deep learning-based lesion segmentations, e.g. in liver MR images or spinal CT images, our experiments showed similar or in some respects superior segmentation quality. Overall, our automatic approach can provide almost expert-like segmentation accuracy in this challenging and ambitious task.


\end{abstract}

\section{Introduction}
\label{intro}

\begin{figure*}[!t]
\centering
\includegraphics[width=\textwidth]{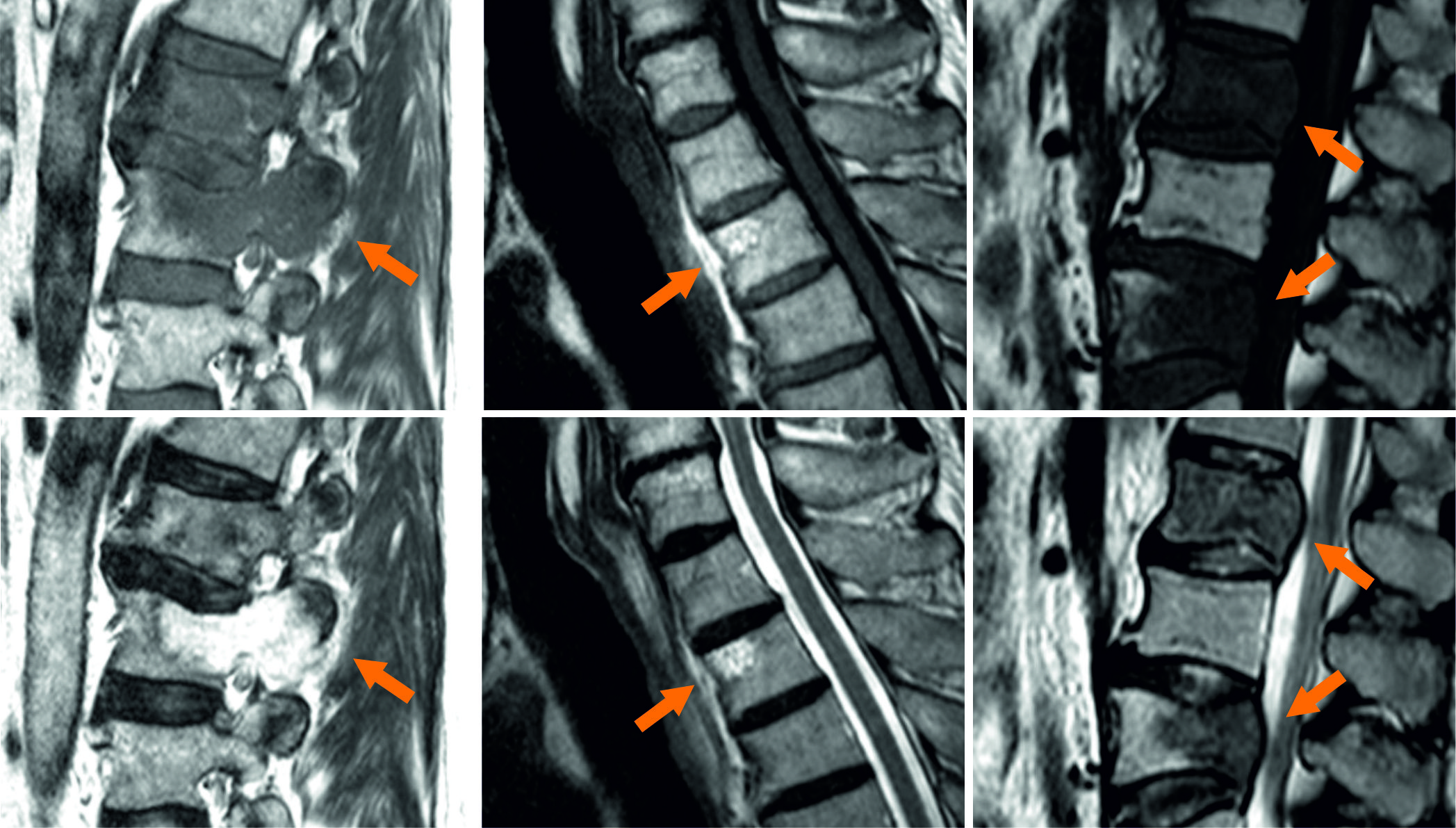}\\[-0.1pt]
\caption{Examples of the shape and appearance variablility of different lesion types (orange arrows) in  $T_1$-weighted (upper row) and within $T_2$-weighted MRI sequences (bottom row). Depicted is a epidural metastasis with an osteolytic vertebral body lesion (left column). The sagittal $T_1$-weighted MR image shows hypointensity with a paraspinal mass, while the  $T_2$-weigthed image displays the lesion hyperintensely. The mid column displays hyperintensitiy in all acquired MR images, which is typical for benign hemangioma. Sclerotic metastases are displayed in the right column, showing characteristic hypointense
signals compared to bone marrow in both MR images.}
\label{fig:metastases}
\end{figure*}

\begin{figure*}[!t]
\centering
\includegraphics[width=\textwidth]{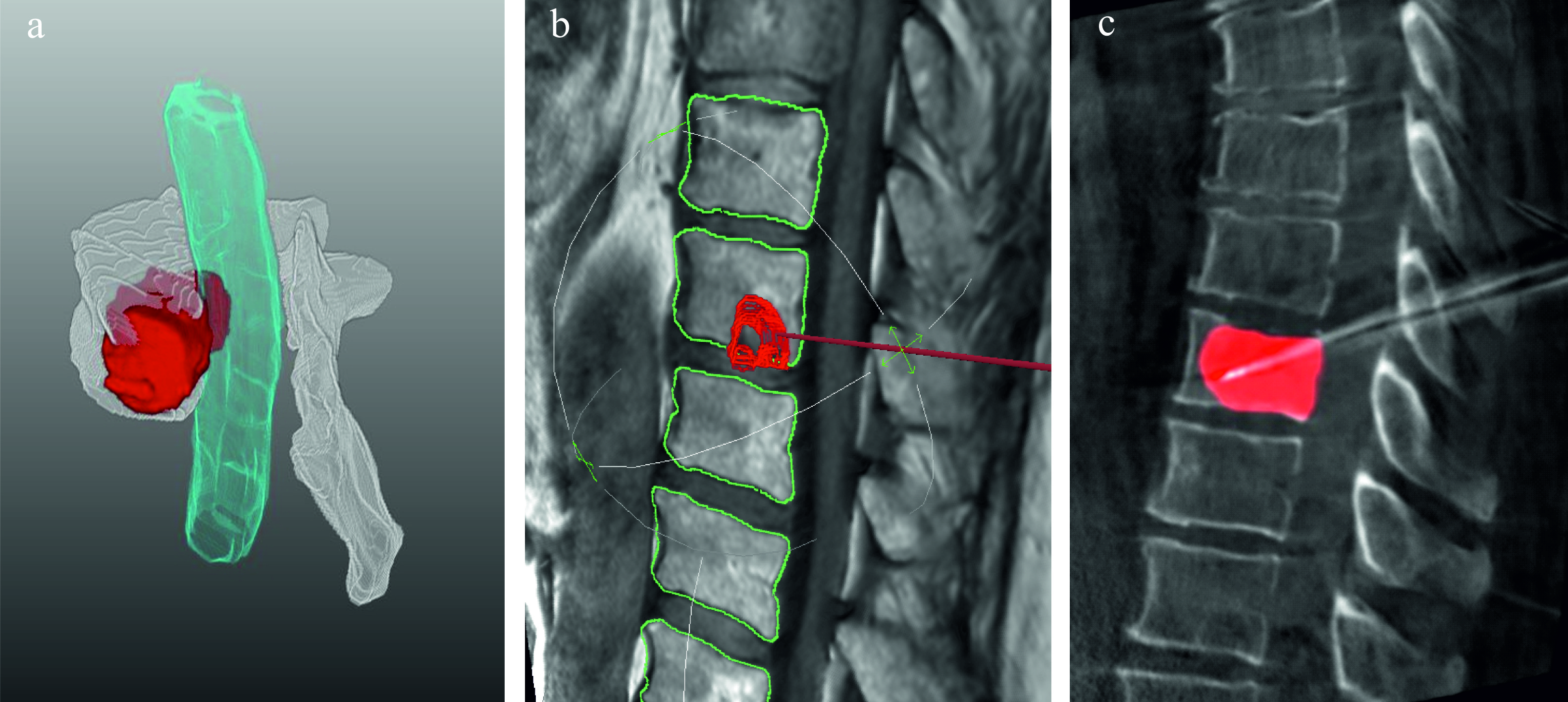}\\[-0.1pt]
\caption{Segmentation of spinal metastases could support multiple procedures throughout the therapeutic workflow: intuitive visualisation of spatial relations to risk structures (a) and applicator pathway optimization (b) while intervention planning, as well as navigation support and visual target zone enhancement during the intervention (c).}
\label{fig:mot}
\end{figure*}

Due to the improvement of medical treatment and diagnostic procedures, life expectancy has increased steadily over the last decades. However, this lifetime gain promotes also age-related diseases like cardiovascular diseases, as well as cancer and cancer induced malicious metastases. The survival time of most malicious carcinomata has increased with improved diagnosis and treatment, hence, the probability to develop metastases raises. Beside liver and lung, bone metastases are the third most likely and thereof up to two thirds are located in the spine \cite{harrington1986metastatic,wong1990spinal}. Spinal metastases can tremendously affect the quality of life by evoking vigorous pain by fractures, bruises, spinal cord and nerve root compressions or neurologic deficits \cite{klimo2004surgical}. Diagnosis and therapy planning can be done with multiple radiological imaging techniques, e.g. planar X-ray radiography, computed tomography (CT), single photon emission computed tomography (SPECT) or magnetic resonance imaging (MRI). The latter overcomes problems with radiation exposure of the aforementioned imaging techniques and has enhanced soft tissue contrast, which promotes early lesion detection and advanced diagnostic performance in terms of osseous lesions. Futhermore, spatial relationships of the metastazised vertebrae and surrounding tissues like the spinal cord or inter-vertebral discs are better visualized in MR than in CT or SPECT imaging. \
Dependending on their origin, there are two common types of bone metastases: lytic lesions, which lead to increased osseous tissue disruption due to further osteoclastic activity and sclerotic lesions, leading to increased osteoblastic activity and therefore, bone tissue production. The specific type tremendously affects the appearance of the metastases in the respective MR imaging sequences, ranging from hypo- to hyperintense image signals compared to non-pathologic vertebral bone structures (see Fig.~\ref{fig:metastases}), making this a highly challenging and ambitious task. \\[1pt]
\noindent With regard to minimally invasive image-guided interventions like radiofrequency ablation (RFA), the segmentation of vertebral metastases constitutes an essential prerequisite for various workflow steps, starting from therapy planning, over the intervention procedure, up to the treatment outcome validation (see Fig. \ref{fig:mot}). First of all, it enables a detailed assessment of extent, shape and spatial relations of the metastases with surrounding risk structures and thus supports the planning of access pathways and positioning of minimally invasive applicators. In addition, patient-individual simulations of RFAs, i.e. ablation zone predictions are based on these segmentations \cite{weihusen2010towards,kroger2006numerical}. Pre-interventionally produced segmentation masks used as overlays onto the intra-operative images can enhance navigation w.r.t. accuracy and time required for metastasis puncturing and therefore, have a beneficial effect on the treatment outcome. Finally, after image registration of pre- and post-operative MRI scans, quantitative outcome validations are made feasible by matching the segmentation masks of metastases and necrosis zones. A computer-assisted approach could save time during routines and recurrent procedures and relieve the workload of radiologists, since segmentation of volumetric image data is time-consuming and fatiguing given the large number of image slices and sequences acquired per patient. \\[1pt]
The main objective of this work was to implement a segmentation approach based on deep Convolutional Neural Networks (CNN) and assess its ability to segment spinal metastases in MR images. Due to its wide-spread applicability in medical image segmentation, we used U-Net-like \cite{ronneberger2015u} networks, with varying architectures and image input configurations to evaluate multiple impacts on the segmentation accuracy. The experiments were carried out with diagnostically acquired MR images of patients who underwent RFAs of spinal metastases.
\\
Beside well established segmentation methods like thres\-hold-based, region-based, classification or model-based approaches, deep learning techniques have been introduced more recently to lesion detection and segmentation tasks. The latter focussed mainly on liver \cite{christ2017automatic,li2017h} and brain \cite{havaei2017brain,kamnitsas2017efficient} lesions, both in CT and MR imaging. 
Segmentation of spinal metastases are highly ambitious, since a variety of anatomical structures with high image contrasts, similar intensities and textures are in close proximity. To this day there are only few publications regarding computer-assisted methods dealing with such spinal lesions and most of them focussed on detection of a specific metastastic type in CT images. Yao et al. \cite{yao2006computer} presented a 2D watershed algorithm to detect potential lytic vertebral lesions with a final classification done by a support vector machine (SVM). Wiese et al. \cite{wiese2012detection} likewise presented an approach for sclerotic spinal metastases detection in CT images with SVMs, but modified this method with graph-cut merging of 2D regions. Wels et al. \cite {wels2012multi} and Hammon et al. \cite {hammon2013automatic} proposed different approaches using multiple random forests discriminative models. 
Roth et al. \cite{roth2015detection} presented the first framework based on a deep CNN, while they used it as a second layer in a two-layered cascade framework to spot candidate lesions for sclerotic spine metastases detection in CT imaging. 
Regarding vertebral metastases detection in MR images, Jerebko et al. \cite{jerebko2007robust} proposed a manually initialized method with simple adaptive thresholding to find candidate lesions. Subsequently, a classification algorithm based on Fisher’s linear discriminant (FLD) analysis was implemented, forced to positively classify at least one candidate in a true lesion, even though multiple false positives were taken into account. 
Wang et al. \cite{wang2017multi} introduced deep CNNs to vertebral metastases detection in MR imaging, by using a Siamese deep neural network (SdNN) approach with multi-resolution analysis and a weighted averaging of neighboring cross-sections to benefit from the similarities and aggregate the detection results. The SdNN comprised three identical multi-layer subnetworks to process each image patch resolution and produced a likelihood map for each MRI slice, where the final classification was done.
In terms of automatic spinal lesion segmentation in CT images, Chmelik et al. \cite{chmelik2018deep} proposed a voxel-wise classification based on a deep CNN with subsequent post-processing to simplify object shapes and produce smooth contours, since the voxel-wise approach usually have scattered object surfaces. \
To summarize, the state-of-the-art regarding computer-assisted methods for spinal lesions in particular focused on the detection, both in CT and MRI images with promising results. Automatic segmentation, however, poses a much greater challenge, which has not been addressed in MR imaging in particular.

\section{Materials and Method}
\label{methods}

\subsection{Image data}

Our dataset contained patient cases who underwent radiofrequency ablations of both, single and multiple spinal metastases, mostly in advanced tumor stages. In total, 40 metastases were assembled for this work, originating from renal cell, prostate, cervical, colon, pancreatic, breast, bladder, stomach, lung, caecal, urothelial and spinocellular carcinoma. For diagnostic and therapy planning purposes, spine MR imaging was performed including sagittal native $T_1$-weighted and $T_2$-weighted MRI sequences. Acquisition settings, e.g. magnetic field strength or repetition time, varied within the dataset. The scan resolution ranged in-plane from $\SI{0.45}{mm}$ to $\SI{1.25}{mm}$ and in-depth from $\SI{3.3}{mm}$ to $\SI{4.8}{mm}$. 
The acquired MRI data was pre-processed by registering cohesive MR sequences patient-wise to the respective $T_1$-weighted image and by cubic interpolation between the original number of sagittal slices (ranged between 15 and 28) to a fixed number of 64 sagittal to provide an almost isotropic spatial resolution. A field expert trained by neuroradiologists manually contoured each metastasis slice-wise, producing a stack of annotated slices per patient or when combined a binary 3D segmentation mask with a defined center point $m_{c}$ in world coordinates.
%augmentation 
Since our dataset was comparably small for a CNN-based approach, we extensively augmented each of the 40 original MRI volumes using the following techniques:

\begin{itemize}
\item \textit{Mirroring:} We applied flips to each patient volume in all directions, although for example vertical, i.e. craniocaudal, flips may appear inappropriate, it had proven to be advantageous for the final results since it prevents fast overfitting. 
\item \textit{Scaling:} We scaled the image volumes with randomly chosen factors between 0.6 and 1.4.
\item \textit{Rotation:} We rotated the image volumes in the range of $\pm \SI{30}{^\circ}$ around the transversal axis and between $\pm \SI{20}{^\circ}$ around the sagittal axis. 
\item \textit{Elastic deformations:} Elastic deformations were applied using random displacement fields with subsequently Gaussian smoothing the grid with a $\sigma$ ranging between $0$ and $0.3$ (cf. \cite{ronneberger2015u}).
\item \textit{Gaussian blur:} A Gaussian filter with $\sigma$ in the range from $0$ to $0.5$ was applied to blur the images.
\item \textit{Gamma transformation:} We applied gamma transformations with $\gamma$ in the range from $0.5$ to $2$ to modify image intensities. 
\item \textit{Translation:} Each patient volume was translated in a range of $\pm \SI{20}{voxels}$ in sagittal and vertical direction w.r.t. the center of the metastasis $m_{c}$ and subsequently cropped to patches of the fixed size of $128 \times 128 \times 64$ voxels.

\end{itemize}

In conclusion, each augmentated and cropped volume was whitened by mean subtraction and a subsequent division by the standard deviation. By excessively using these augmentation techniques we generated 5,250 volumetric training samples per fold respectively 336,000 cross-section samples, if treated as individual slices. It was ensured that each patch contained at least fractions of metastasized tissue.

\subsection{CNN architecture}
\label{CNN architecture}

\begin{figure*}[!t]
\centering
\includegraphics[width=\textwidth]{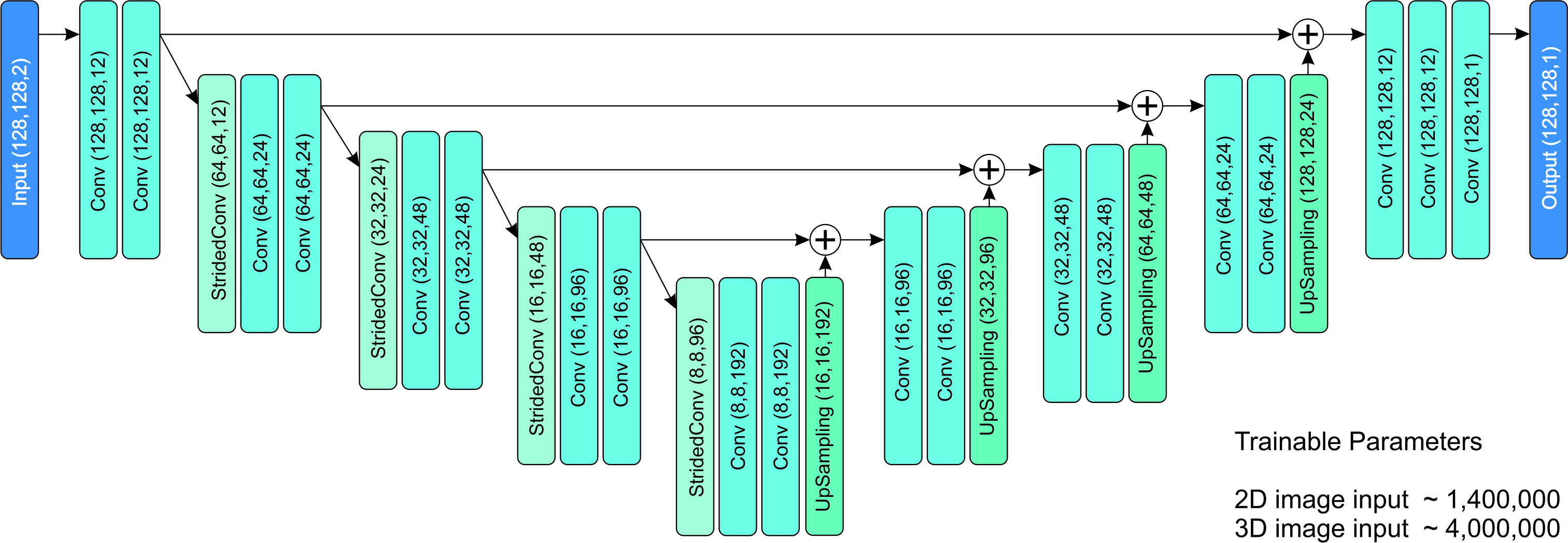}\\[-0.1pt]
\caption{The U-Net structure used for multi-modal 2D image input, with convolutional layers including batch normalisation, strided convolutions for downsampling and upsampling layers. The architecture for three-dimensional input is analogous to the one shown above. A significant difference between the two variants is the number of trainable parameters, which is about 2.85 times higher in the 3D case.}
\label{fig:network-c}
\end{figure*}

With regards to the fact that the commonly used U-Net architecture from Ronneberger et al. \cite{ronneberger2015u} is still state-of-the-art in terms of various medical segmentation tasks \cite{hille2018vertebral}, we implemented a minimally modified U-Net using $Keras$ and $Tensorflow$, which incorporates either 2D or 3D image data. U-Net represents an encoder-decoder architecture well suited for medical images, whose decoder path combines semantic information from the deeper layers of the network with higher resolution feature maps yielded by the encoder via skip connections (see Fig. \ref{fig:network-c}). 
Our network processed 2D patches of size $128 \times 128$ pixels or volumes of size $128 \times 128 \times 64$ voxels for 3D image input. Each convolutional layer was followed by a batch normalisation and had a kernel size of $3 \times 3~(\times 3)$ except the last one, which applies a $1 \times 1~(\times 1)$ kernel to reduce the dimensionality to the desired output size. We used strided convolutions with a stride value of 2 for downsampling our images. Furthermore, we replaced up-convolutions by simplified upsampling layers, which have been found to be equally effective, while being less computationally expensive \cite{isensee2017brain}.  A Rectified Linear Unit (ReLU) was used as the activation function for all convolutional layers, except the last one again, where a sigmoid function was applied to provide values between 0 and 1. Multi-modal image input was incorporated in the most straightforward way, i.e. each MRI sequence was represented by an input channel. \\[1pt]
\noindent The training set of each fold consisted of 63,000 randomly shuffled volumes for 3D input or 4,032,000 randomly shuffled slices for 2D input. We used a single epoch, while the number of iterations was equal to the number of available samples. With the Tversky Loss (TL) as proposed by Salehi et al. \cite{10.1007/978-3-319-67389-9_44}, we used a modified form of the Tversky index \cite{tversky1977features} as a loss function, which is defined as

\begin{equation}
TL(\alpha,\beta) = \frac{2  \sum \limits_{i = 1}^N {(r_{0i} p_{0i})}}{\sum \limits_{i = 1}^N {(r_{0i} p_{0i})} + \alpha \sum \limits_{i = 1}^N {(r_{1i} p_{0i})} + \beta \sum \limits_{i = 1}^N {(r_{0i} p_{1i})} }
\end{equation}  

where $p_{0i}$ is the probability for a voxel i to be a lesion and $p_{1i}$ to be non-lesion. For a lesion voxel $r_{0i}$ is 1 and for a non-lesion voxel $r_{0i}$ is 0, vice verse for $r_{1i}$. The weights $\alpha$ and $\beta$ affect the penalities for false positives and false negatives. 
Furthermore, we used Adam \cite{kingma2014adam} as an optimizer with a starting learning rate of $0.001$ and a mini-batch size of 2 samples for volumetric and 32 for slice-wise input data. Finally, a threshold of $0.5$ was applied to the ouput layer produce binary output images. \

\subsection{Experimental Design}

Using our augmented dataset, we evaluated multiple network configurations in terms of their ability to segment spinal metastases. For this purpose we applied our data either slice-wise or as image volumes to the network described above. We decided to not change the basic architecture for both volumetric and slice-wise input in order to largely exclude further influencing factors, e.g. by varying layer or channel numbers. Additionally, we subdivided the experiments according to the used input modalities, i.e. single or combined MR sequences. Our training design consisted of stratified 8-fold cross-validation over disjunct subsets of five patients per validation set. The stated results represent the average of all 8 cross-validation folds. Due to the small amount of available data and since we did not base any training and design decisions on intermediate validation results (no look-ahead bias), we decided to refrain from a separate test set, as it would result in too few samples for a promising training.     

\section{Results}
\label{sesults}

\subsection{Evaluation}

Expertly annotated lesion segmentations were produced considering co-reg\-istered MR sequences of each patient within a synchronized viewer. The network described in Section \ref{CNN architecture} was applied to the image data. To quantify our results, we used Dice similarity coefficients to measure the percentage of volume overlap, as well as voxel-wise sensitivity (true positive rate, TPR) and specificity (true negative rate, TNR), since some of the related work used both as quality measurements. The above mentioned are defined as follows:

\begin{equation}
{\textstyle Dice = \frac{2~|R_1 \cap P_1|}{|R_1| + |P_1|}, \quad TPR = \frac{|R_1 \cap P_1|}{|R_1|}, \quad TNR = \frac{|R_0 \cap P_0|}{|R_0|}}
\end{equation}

with $R_1$ and $P_1$ as foreground voxels of reference and prediction and analogously, $R_0$ and $P_0$ as the corresponding background voxels. However, sensitivity and specificity are not commonly used to evaluate medical image segmentations, since they are highly sensitive to segment's size \cite{taha2015metrics}. The given results were generated exclusively on patient volumes, even if the segmentations with 2D input were predicted slice-wise. Thus, the 2D predictions were merged patient-wise.

\subsection{Experimental Results}

In order to reasonably classify the achieved results, it is advantageous to investigate the inter-reader variability to get an impression of the accuracy and deviation, if two expertly annotated segmentations of a specific metastasis were matched. Therefore, we produced a second ground truth of a randomly chosen subset of 15 metastases by a neuroradiologist and matched those with the corresponding primary segmentation masks. Again, we determined Dice scores, as well as sensitivity and specificity. The computed Dice values per metastasis ranged from $\SI{70.4}{\%}$ to $\SI{89.7}{\%}$ with an average of $79.4 \pm \SI{6.2}{\%}$. Mean sensitivity was $80.0 \pm \SI{11.9}{\%}$ and the average specificity was $98.2 \pm \SI{2.1}{\%}$. The Dice coefficient scores as well as sensitivity and specificity of automatic and expertly annotated segmentations of all performed experiments can be found in table~\ref{tab:dice_results}. \

\begin{table*}[htbp]
  \centering
  \caption{Experimental results for each input configuration depending on the used modalities ($T_1$-, $T_2$-weighted MRI sequences), as well as a slice-wise (2D) or volume (3D) processing. The scores of the inter-reader variability (IRV) are listed in the last column.}
\resizebox{\textwidth}{!}{%
\begin{tabular}{ccccc ccccccc}

\multicolumn{8}{c}{ }\\
\toprule
& & \multicolumn{3}{c}{2D} & & \multicolumn{3}{c}{3D} & & \multirow{2}{*}{IRV}\\
&  & $T_1$ & $T_2$ & $[T_1+T_2]$ & & $T_1$ & $T_2$ & $[T_1+T_2]$ \\
\midrule
%& \\
\multirow{2}{*}{$Dice~[\%]$} & mean & 77.4 & 65.4 & 77.6 & & 73.7 & 64.7 & 74.8 &  & 79.4\\
& std & 12.4 & 21.7 & 10.8 & & 15.6 & 20.8 & 13.6 &  & 6.2\\
\multirow{2}{*}{$Sensitivity~[\%]$} & mean & 76.2 & 71.9 & 78.9 & & 71.9 & 65.2 & 73.7 & & 80.0\\
& std & 17.4 & 21.6 & 15.8 & & 20.2 & 24.51 & 18.6 & & 11.9\\
\multirow{2}{*}{$Specificity~[\%]$} & mean & 98.5 & 97.5 & 98.4 & & 98.4 & 97.85 & 98.33 & & 98.2\\
& std & 2.0 & 3.3 & 1.9 & & 2.1 & 2.7 & 2.1 & & 2.1\\
\bottomrule      
\end{tabular}}%
\label{tab:dice_results}
\end{table*}

\section{Discussion}

\begin{figure*}[!t]
\centering
\includegraphics[width=\textwidth]{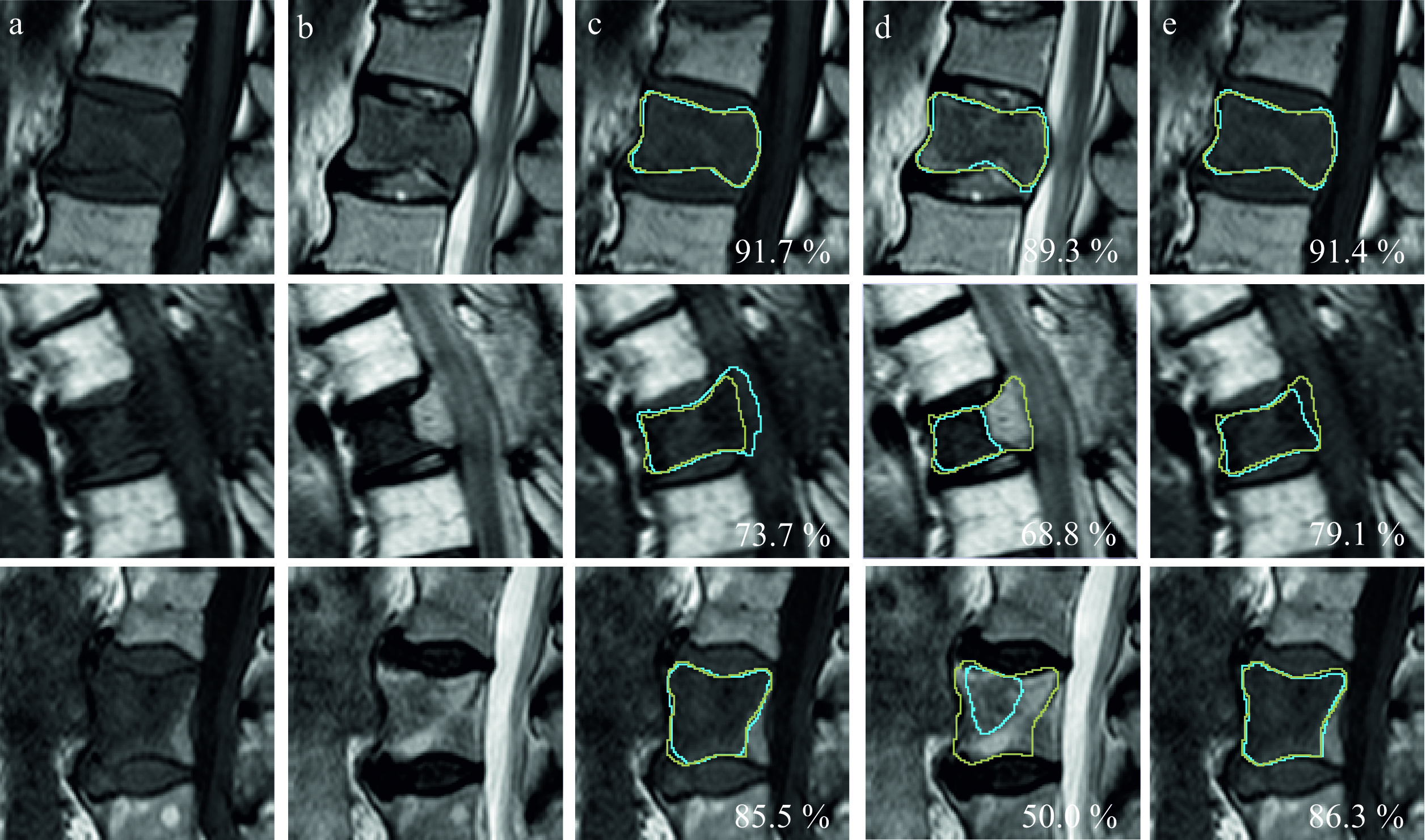}\\[-0.1pt]
\caption{Comparison of the expertly annotated data (green contours) with our automatically produced segmentations (blue contours) with 2D image input for three exemplary cases. Dice scores indicate the segmentation accuracy for each patient case and network configuration and are stated in the lower right corners. From left to right: (a) original $T_1$-weighted MRI sequence, (b) original $T_2$-weighted MRI sequence, (c) result with only $T_1$-weighted image data, (d) result with only $T_2$-weighted image data, (e) result with combined $T_1$- and $T_2$-weighted image data.}
\label{fig:results_im}
\end{figure*}

\begin{figure*}[!t]
\centering
\includegraphics[width=1\textwidth]{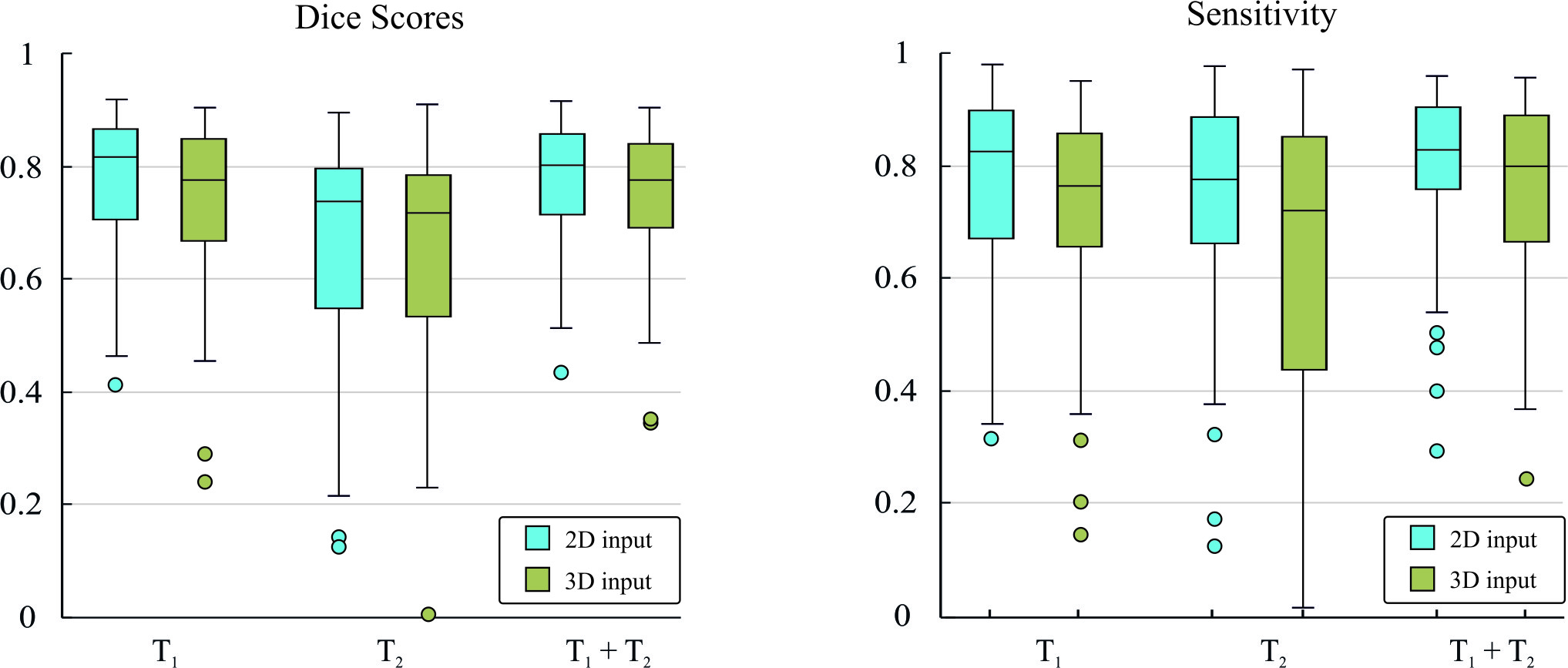}\\[-0.1pt]
\caption{Measured Dice scores and sensitivity depending on the imaging modality and input dimension. Box edges mark the 25th and 75th percentiles, the central box line marks the median value and whisker the most extreme values not considered outliers.}
\label{fig:results_modal}
\end{figure*}

Our experimental results have been matched with the inter-reader variability as well as with relevant state-of-the-art results in literature. To start with, some representative cases are shown in Figure \ref{fig:results_im}, depicting satisfactory results as well as challenging cases with exemplary inaccuracies. Among the latter, false positively classified voxels, mostly in the adjacent tissue of the vertebral bodies and inaccurate segmentations at the transverse and spinous processes are the most common. Segmentation tasks in this particular area are highly ambitious, since a variety of anatomical structures with high image contrasts, similar intensities and textures are in close proximity. This even hampers manual contouring by experts, which is reflected in the relatively low Dice scores of the inter-reader variability. Furthermore, exceptionally shaped metastases, especially if they were not roughly star-convex, or metastases with highly heterogenous image signals represent a challenging task for automatic approaches like CNNs. Here it is particularly troublesome if the training set does not represent sufficient variance of the real data. This is indicated, for example, by the comparatively high standard distribution. \\[1pt]
\noindent Regarding the input modalities, it was found that $T_1$-weighted images were the most valuable, since the best results could be achieved in experiments either with $T_1$-weighted data alone or if it was part of multimodal image input (see Fig. \ref{fig:results_modal}). This can be attributed to the predominantly uniform appearance of the metastases in $T_1$-weighted images, whether lytic or sclerotic, which are hypointense compared to surrounding bony structures. Combining $T_1$-weighted images with $T_2$-weighted MR data hardly showed any significant differences in the mean accuracy compared to an input of exclusive $T_1$-weighted images ($\SI{77.6}{\%}$ vs. $\SI{77.4}{\%}$), although the standard deviation could be reduced markedly ($\SI{12.4}{\%}$ to $\SI{10.8}{\%}$).  
The experiments using solely $T_2$-weighted images yielded the worst results most likely due to the fact that lytic and sclerotic lesions most notably differ in this particular imaging sequence and may present conflicting information to our CNNs, if in the same training set. Hence, $T_2$-weighted images rather support and improve the robustness in combination with $T_1$-weighted input than yield satisfactory results themselves.
The above drawn conclusions regarding imaging sequences also hold true for volumetric input. In general, the achieved Dice scores with 2D input were on average $\SI{2.4}{\%}$ higher than with 3D input, which could be attributed to the increased complexity and number of trainable parameters (1,400,000 vs. 4,000,000) to be optimized when extending the network from 2D to 3D input. With respect to input sequences and dimension, the sensitivity and specificity validation led to similar conclusions analogous to the Dice scores (see Tab. \ref{tab:dice_results}). In order to classify the achieved results of our experiments in their general segmentation quality, it is convenient to compare them with the inter-reader variability of manually produced expert segmentations, which could be seen as an indicator for the complexity and challenge of such a segmentation task. With on average $\SI{77.6}{\%}$, our best segmentation results were close to the mean inter-reader variability ($\SI{79.4}{\%}$), although the standard deviation was significantly higher ($\SI{10.8}{\%}$ vs. $\SI{6.2}{\%}$). Therefore, we can conclude that our results are largely on par with expertly annotated segmentations, but still lacking the robustness of trained experts with regard to individual patient cases. \\[1pt]
\noindent It is difficult to compare our results with related works, since there are, to our best knowledge, no studies regarding automatic spinal metastases segmentation in MR imaging. Thus, comparison is indirect and refers to CNN-based segmentation approaches for instance of liver and brain lesions, as well as a recently published work by Chmelik et al. \cite{chmelik2018deep} for spinal CT data.
Depending on the used data sets, CNN-based brain tumor segmentations in MR images achieved Dice scores up to $\SI{88}{\%}$ \cite{havaei2017brain} (brain tumor segmentation challenge 2013) or $\SI{84.9}{\%}$ \cite{kamnitsas2017efficient} (brain tumor segmentation challenge 2015). Segmentation of liver lesions in MR images achieved Dice coefficients of $\SI{69.7}{\%}$ \cite{christ2017automatic}, in CT images up to $\SI{72.2}{\%}$ \cite{li2017h}. Overall, our results are comparable with the segmentation accuracies of liver lesions, although our data base was comparatively small (40 patient cases vs. 200 cases in \cite{li2017h}), which is also reflected in the fairly high standard deviation of our results. 
Chmelik et al. \cite{chmelik2018deep} were one of the first to adapt a CNN to vertebral metastases segmentation in CT images. They achieved a voxel-wise sensitivity rate of $\SI{74}{\%}$ for sclerotic and $\SI{71}{\%}$ for lytic lesions as well as a specificity rate of $\SI{88}{\%}$ (sclerotic) and $\SI{82}{\%}$ (lytic). 
In comparison, our results including $T_1$-weighted images are somewhat better (mean sensitivity of $\SI{77.6}{\%}$), though the experiments only with $T_2$-weighted MR data clearly lack accuracy. Additionally, it is important to account for the differences in spatial resolution (slice thickness of $\SI{0.67}{mm}$ vs. our average $\SI{3.50}{mm}$) and the effects of high spatial anisotropy and therefore, partial volume effects, which could hamper automatic segmentation approaches.\\

\section{Conclusion}
\label{concl}

Automatic spinal metastases segmentation has the potential to support therapy planning, performance and treatment outcome validation of minimally invasive interventions such as RF ablations. We presented a CNN-based segmentation approach for spinal metastases in diagnostic MR images. We assembled a dataset including metastases of both, lytic and sclerotic type and examined the impact of various input modalities and dimensions on the segmentation accuracy. Our experimental results have been quantitatively compared to the inter-reader variability and results in literature, although the latter focussed on other metastazised organs or imaging modalities. Due to the absence of directly comparable works, the inter-reader variability most likely indicates the quality of our achieved results. With $\SI{77.6}{\%}$ the best result of our automatically performed segmentation is quite on par with the inter-reader variability ($\SI{79.4}{\%}$), indicating reasonably accurate and almost expert-level segmentation quality. Accordingly, our proposed work constitutes a promising approach towards this ambitious and challenging issue.\\

%\begin{acknowledgements}
%\end{acknowledgements}

\section{Compliance with Ethical Standards}

\noindent \textbf{Funding}
This study was partly funded by the Federal Ministry of Education and Research within the Forschungscampus \textit{STIMULATE} (grant number \newline 13GW0095A).\\

\noindent \textbf{Conflict of interest}
 The authors declare that they have no conflict of interest.\\

\noindent \textbf{Ethical approval} All procedures performed in studies involving human participants were in accordance with the ethical standards of the institutional and/or national research committee and with the 1964 Helsinki declaration and its later amendments or comparable ethical standards.\\

\noindent \textbf{Informed consent} Informed consent was obtained from all individual participants included in the study.

\bibliographystyle{spmpsci}      % mathematics and physical sciences
\bibliography{biblio}   % name your BibTeX data base

\end{document}